\pdfoutput=1
\documentclass{nice-article}

\usepackage{breakcites}
\usepackage{ebproof}
\usepackage{macros}
\Crefname{diagram}{Diagram}{Diagrams}
\Crefname{identity}{Identity}{Identities}

\NewDocumentCommand\T{g}{\uptextbf{T}\IfValueT{#1}{\prn{#1}}}
\NewDocumentCommand\Ar{m}{\mathds{#1}}
\NewDocumentCommand\Shape{d<>g}{\mathsf{S}\IfValueT{#1}{\Sub{#1}}\IfValueT{#2}{\prn{#2}}}
\NewDocumentCommand\Eqns{g}{\mathsf{E}\IfValueT{#1}{\prn{#1}}}
\NewDocumentCommand\TyBool{}{\mathsf{bool}}
\NewDocumentCommand\TmTrue{}{\mathsf{tt}}
\NewDocumentCommand\TmFalse{}{\mathsf{ff}}
\NewDocumentCommand\Code{m}{\texttt{#1}}
\NewDocumentCommand\CodeObjTerm{}{\Code{T}}
\NewDocumentCommand\CodeArrTerm{m}{\Code{t}\Sub{#1}}

\hypersetup{
  pdfauthor = {Daniel Gratzer and Jonathan Sterling},
  pdftitle = {Syntactic categories for dependent type theory: sketching and adequacy},
}

\title{%
  \normalfont
  \LARGE
  Syntactic categories for dependent type theory:\\[4pt]
  sketching and adequacy
}%
\author{Daniel Gratzer\and Jonathan Sterling}

\begin{document}

\maketitle

\begin{abstract}
  We argue that locally Cartesian closed categories form a suitable doctrine
  for defining dependent type theories, including non-extensional ones. Using
  the theory of sketches~\citep{kinoshita-power-takeyama:1999}, one may define
  syntactic categories for type theories in a style that resembles the use of
  Martin-L\"of's Logical Framework~\citep{nordstrom-peterson-smith:1990},
  following the ``judgments as types''
  principle~\citep{harper-honsell-plotkin:1993,martin-lof:1987:wgl}.

  The concentration of type theories into their locally Cartesian closed
  categories of judgments is particularly convenient for proving syntactic
  metatheorems by semantic means (canonicity, normalization, \emph{etc.}).
  Perhaps surprisingly, the notion of a \emph{context} plays no role in the
  definitions of type theories in this sense, but the structure of a class of
  display maps can be imposed on a theory \emph{post facto} wherever needed, as
  advocated by the Edinburgh school and realized by the
  \texttt{\%worlds} declarations of the Twelf proof
  assistant~\citep{harper-honsell-plotkin:1993,pfenning-schuermann:1999,harper-licata:2007}.

  \Citet{uemura:2019} has proposed representable map categories together with a
  stratified logical framework for similar purposes. The stratification in
  Uemura's framework restricts the use of dependent products to be strictly
  positive, in contrast to the tradition of Martin-L\"of's logical
  framework~\citep{martin-lof:1987:wgl,nordstrom-peterson-smith:1990} and
  Schroeder-Heister's analysis of higher-level
  deductions~\citep{schroeder-heister:1987}.

  We prove a semantic adequacy result for locally Cartesian closed categories
  relative to Uemura's representable map categories: if a theory is definable
  in the framework of Uemura, the locally Cartesian closed category that it
  generates is a conservative (fully faithful) extension of its syntactic
  representable map category. On this basis, we argue for the use of locally
  Cartesian closed categories as a simpler alternative to Uemura's
  representable map categories.
\end{abstract}

\section{Introduction}\label{sec:intro}

\begin{node}\label{node:what-is-dtt}
  What kind of objects are dependent type theories and their models?
  Unfortunately there are many possible answers to this question:
  \begin{enumerate}

    \item \emph{Comprehension categories}~\citep{jacobs:1999} express the
      structure of a category of contexts equipped with separate notions of
      type and term, connected by a context extension operation.  Sometimes a
      comprehension category is ``split'' (modeling strictly associative
      substitution).

    \item \emph{Categories with attributes}~\citep{cartmell:1978} are
      \emph{full} split comprehension categories; fullness means that the
      notion of an element is \emph{derived} from the context extension.

    \item \emph{Categories with families}~\citep{dybjer:1996} express the
      notion of type, term, and context extension as a special kind of
      \emph{universe} in the category of presheaves over a category of
      contexts. These are easily seen to be equivalent to categories with
      attributes, but they are arguably more type theoretic in style.

      \citet{awodey:2018:natural-models,fiore:2012} have independently
      reformulated categories with families in terms of \emph{representable
      natural transformations}; Awodey has coined the name \emph{natural model}
      for this formulation.

    \item \emph{Contextual categories}~\citep{cartmell:1978} or
      \emph{C-systems}~\citep{voevodsky:2015:c-system} are categories with
      attributes together with a structure that equips each context with a
      ``length'', reflecting the inductive generation of contexts in some
      presentations of the raw syntax of type theory.

    \item \emph{Display map categories}~\citep{taylor:1986,taylor:1999} or
      \emph{clans}~\citep{joyal:2017} express the data of a category of
      contexts equipped with a class of ``display maps'' that generalize
      context extensions.

  \end{enumerate}
\end{node}

\begin{node}
  The notion of a \emph{contextual category} or a \emph{C-system} is not as
  useful as it might at first seem: contextual categories differ from
  categories with attributes only by elevating to the status of a definition
  the incidental aspect of certain raw syntax presentations of type theory that
  contexts have a length. This influence of (raw) syntax on semantics has so
  far imparted no practical leverage, even when proving metatheorems of a
  syntactical nature: indeed, no theorem of dependent type theory could ever
  have depended on the fact that not every context is of length \emph{one}.

  Categories with attributes and natural models are in essence the same notion.
  Comprehension categories on the other hand occupy an awkward position: full
  split comprehension categories are the same as categories with attributes and
  natural models, and full non-split comprehension categories are the same as
  clans.
  From our perspective, the \emph{canonical} notions among the above are
  therefore clans and natural models, representing the weak and strict notions
  of dependent type theory respectively.
\end{node}

\begin{node}
  We have enumerated a number of scientific hypotheses as to what a model of
  type theory ought to be; aside from clans, however, \cref{node:what-is-dtt}
  does not pose a definition of the \emph{syntactic/classifying category} of a
  given type theory in the sense of functorial
  semantics~\citep{lawvere:thesis}. Recently \citet{uemura:2019} has proposed
  \emph{representable map categories} to serve as the syntactic categories that
  classify the natural models of a given type theory, which we discuss below in
  \cref{sec:rmcats}.
\end{node}

\NewDocumentCommand\REP{m}{{#1}\Sup{\mathit{rep}}}

\subsection{Uemura's representable map categories}\label{sec:rmcats}

\begin{node}
  A representable map category is a finitely complete category $\RepThy$
  equipped with a pullback stable subcategory $\REP{\RepThy}\subseteq\RepThy$
  of ``representable maps'' such that $\RepThy$ contains dependent products
  along representable maps. Representable maps correspond roughly to display
  maps, and an object $\Gamma:\RepThy$ such that
  $\Mor{\Gamma}{\ObjTerm{\RepThy}}$ is representable can be thought of as a
  \emph{context}. An arbitrary (non-representable) object of $\RepThy$ stands for
  a \emph{judgment} --- something that will be taken to a not necessarily
  representable presheaf in the natural model semantics.
\end{node}

\begin{node}
  While a clan expresses the categorical structure of the category of contexts
  of a given type theory, a representable map category attempts to express the
  categorical structure of its \emph{category of judgments}. In particular,
  while a clan need not be finitely complete (the diagonal is not a display map
  except in extensional type theories), unrestricted pullback in representable
  map categories corresponds to the fact that type theory has \emph{judgmental
  equality}.
  Likewise, the presence of (non-representable) dependent products along
  representable maps corresponds exactly to the \emph{hypothetico-general
  judgment} of dependent type theory in the sense of \citet{martin-lof:1996}.
\end{node}

\begin{node}
  The fact that dependent products need exist only along representable maps
  corresponds to the way that dependent type theory is conventionally presented
  using hypothetical judgments of one level only (corresponding by transpose to
  context extension). This realistic stratification, however, is not at all
  forced: Martin-L\"of himself has promoted a presentation of the syntax of
  dependent type theory that supports hypothetical judgments of arbitrary
  level~\citep{nordstrom-peterson-smith:1990,martin-lof:1987:wgl,schroeder-heister:1987}.

  One commonly cited example of the use of higher-level judgments is to present
  dependent products in terms of a ``fun-split''
  operator~\citep{nordstrom-peterson-smith:1990}, but this example is not so
  convincing considering that the presentation is strictly isomorphic to one
  not involving a higher-level judgment.\footnote{\emph{Pace} the observation
  of \citet{garner:2009} that the fun-split formulation is strictly stronger
  than the conventional formulation \emph{in the absence of the $\eta$-law} ---
  perhaps a more realistic title for the cited work would have been ``On the
  strength of dependent \emph{non-products} in the type theory of
  Martin-Löf''.} A more convincing example is furnished by the W-type, whose
  elimination rule apparently cannot even be written down without higher-level
  hypothetical judgments in the absence of dependent product types.
\end{node}

\subsection{Locally Cartesian closed categories}

\begin{node}
  If Uemura's notion of representable map category aims to capture the
  restriction of a category of judgments to contain just the hypothetical
  judgments tracked by context extensions, it is reasonable to consider the
  categorical structure of judgments \emph{absent} such a stratification. Of
  course, as soon as we have both judgmental equality and unrestricted
  hypothetical judgment, the category of judgments is nothing less than a
  locally Cartesian closed category.

  Such a syntactic category may be equipped with a class of display maps, but
  this class is no longer intertwined with the definition of the syntactic
  category. In contrast, one cannot even write down the closure of the type
  theory \emph{qua} \citet{uemura:2019} under \emph{(e.g.)} function types unless
  certain maps are representable.
\end{node}

\begin{node}
  Because the category of judgments of a given type theory can be defined
  independently of the structure of a class of display maps / notion of
  context, it is in fact reasonable to avoid imposing such a structure
  until it is needed. The need for a notion of context arises in several situations:
  \begin{enumerate}

    \item Tautologically, one needs a notion of context when defining a formal
      ``gammas and turnstiles'' presentation of a given type theory.

    \item To correctly state results like decidability of judgmental equality,
      one also needs a notion of context: judgmental equality can only
      decidable relative to a class of display maps that \emph{does not
      include} all diagonals~\citep{castellan-clairambault-dybjer:2017}.

    \item A class of display maps can be used to present an
      $\infty$-categorical structure, as in Joyal's clans~\citep{joyal:2017} and
      Lurie's pre-geometries~\citep{lurie:dag:5}.
  \end{enumerate}

  The idea of identifying the relevant display maps separately from a theory
  and locally to a given construction was promoted and used to great effect by
  the exponents of the Edinburgh school of logical
  frameworks~\citep{harper-honsell-plotkin:1993,harper-licata:2007}; the
  function of the \texttt{\%worlds} declarations of the Twelf proof
  assistant~\citep{pfenning-schuermann:1999} is nothing more than to specify
  the class of contexts relative to which a given metatheorem should hold,
  since almost no metatheorems hold unconditionally.
\end{node}

\begin{node}
  While the presence of higher-level hypothetical judgments in the language of
  locally Cartesian closed categories is convenient, it is \emph{a priori}
  correct to worry about whether it evinces an exotic (and therefore
  inadequate) notion of syntax for a conventional type theory. In
  \cref{sec:conservative}, we prove that the presence of hypothetical judgments
  of arbitrary level is a \emph{conservative} extension of the stratified
  language of Uemura's representable map categories; our result can be seen as
  a \textbf{semantic adequacy theorem} for a encodings of type theories as locally
  Cartesian closed categories.
\end{node}

\subsection{Presenting syntactic categories by generators and relations}

\begin{node}
  While it is very simple to describe how an algebraic theory can be built from
  generators and relations, it is more technical to do so for the syntactic
  category of a dependent type theory. In essence this is because the
  collection of derived generators of a given dependent type theory appears
  \emph{a priori} to depend on all the relations of that theory, a difficulty that
  can be tracked to the presence of the \emph{conversion rule}:
  \begin{prooftree*}
    \hypo{\Gamma\vdash M : A}
    \hypo{\Gamma\vdash A \equiv B\ \textit{type}}
    \infer2{
      \Gamma\vdash M : B
    }
  \end{prooftree*}

  Moreover, the sort of a given generator may need to mention a pullback or
  dependent product that involves some other generators. Hence it is not
  immediately clear how to define all the generators before freely adding
  \emph{(e.g.)} dependent products.
\end{node}

\begin{node}\label{node:logical-frameworks}
  \emph{Logical frameworks.}
  One approach to ``tie the knot'' between generators, relations, and derived
  structure (pullback, dependent products, \emph{etc.}) is to use a
  \emph{logical framework}, in which these notions are all interleaved; this is the approach of
  \citet{cartmell:1978,nordstrom-peterson-smith:1990,uemura:2019,harper:2020:slf}.\footnote{\citet{harper-honsell-plotkin:1993}
  also define a logical framework, but this logical framework only supports
  ``pure'' theories with only generators and no relations.} Logical frameworks
  are very user-friendly, but it can be somewhat technical to make the
  connection between the syntactical artifacts of a logical framework and the
  categorical/mathematical objects they are meant to present.
\end{node}

\begin{node}\label{node:sketches}
  \emph{Sketches.}
  A less syntactic and vastly more general way to present almost any kind of theory by generators and
  relations is furnished by the language of \emph{sketches} for a given (finitary) 2-monad
  or ``doctrine'':
  \begin{enumerate}

    \item In the doctrine of categories with finite products, sketches present
      algebraic theories.

    \item In the doctrine of categories with finite limits, sketches present
      essentially algebraic theories.

    \item In the doctrine of locally Cartesian closed categories, sketches
      present (the categories of judgments of) dependent type theories.

  \end{enumerate}

  Sketches address the apparent interleaving between generators, relations, and
  derived structure in a different way from logical frameworks
  \cref{node:logical-frameworks}. The generators are defined all at once, prior
  to imposing relations or adding derived structure; if a generator in a finite
  limit sketch needs to mention \emph{(e.g.)} a pullback, this object is added
  \emph{formally} as an additional generator and then ``marked'' as something
  that needs to become a real pullback when the sketch is knitted together into
  a finitely complete category. This picture works for any kind of theory that
  can be defined as a 2-monad!
\end{node}

\begin{node}
  The benefit of 2-monads and their sketches is that they are considerably more
  general than any specific logical framework, and they likely encompass the
  expressivity of almost any future logical framework.  Moreover, the
  connection between a sketch and the theory it presents can be rigorously
  developed once and for all without needing to deal with the purely
  bureaucratic aspects of syntax (\emph{e.g.}\ variable binding, conversion,
  presupposition lemmas, \emph{etc.}) that have engrossed and preoccupied some
  parts of the type theoretic research community for a number of years. Once a scientist understands
  the language of sketches, there is nothing to stop her from using a logical
  framework as a convenient \emph{notation}, keeping in mind the
  ``compilation'' to the more basic notion.
\end{node}

\begin{node}
  For the sake of exposition, we recall the basics of 2-monad theory and sketches from
  \citet{kinoshita-power-takeyama:1999} in \cref{sec:doctrines,sec:sketching}. In
  \cref{sec:sketching:examples} we argue for the use of locally Cartesian closed categories as a
  convenient way to present the syntactic categories of (strict) dependent type theories. Only the
  conservativity result of \cref{sec:conservative} is novel.
\end{node}

\section{Doctrines and the theory of 2-monads}\label{sec:doctrines}

\begin{node}
  There are many different kinds of theory, each corresponding to different
  kinds of categorical structure: for instance, the structure of finite
  products, finite limits, finite colimits, infinite colimits, exponentials,
  etc.\ all give rise to different ``notions of theory'' or \emph{doctrines}.
  The intuitive idea of a doctrine is realized at a technical level by the
  notion of a 2-monad, exposed in \cref{sec:2-monads} below.
\end{node}

\begin{node}
  In this section we recall the basic 2-monad theory necessary for this note. We emphasize that a
  deep understanding of this material not a prerequisite for sketching type theories; a casual
  reader may take \cref{node:lcccs-are-monadic} as given and proceed to
  \cref{sec:sketching}.
\end{node}

\subsection{The theory of 2-monads}\label{sec:2-monads}

\begin{node}
  A strict 2-monad $\T$ on $\CAT$ (resp. $\CAT[g]$) is an ordinary monad on
  $\CAT$ regarded as a 1-category, admitting an enrichment over
  $\CAT$ (resp. $\GRPD$).\footnote{Note that $\CAT$ or $\GRPD$ enrichment of a
  monad (indeed, an endofunctor) is merely a property, not a
  structure~\citep{power:2011}.}
  The enrichment means that $\T$ must not only assign to each functor
  $\Mor[f]{\CCat}{\DCat}$ a functor $\Mor[\T{f}]{\T{\CCat}}{\T{\DCat}}$: it also
  assigns to each 2-cell $\Mor[\alpha]{f}{g}$ a 2-cell
  $\Mor[\T{\alpha}]{\T{f}}{\T{g}}$, \emph{strictly} respecting identity and
  composition.
\end{node}

\begin{node}
  While we will not be overly pedantic about the details, we must discuss the
  notion of a monad algebra for a 2-monad $\T$. A monad algebra is a category
  $\ACat$ together with a morphism $\Mor[a]{\T{\ACat}}{\ACat}$ which satisfies
  the following two equations, familiar from the 1-categorical case:
  \begin{gather*}
    \begin{tikzpicture}[diagram,baseline = (A'.base)]
      \node (A) {$\ACat$};
      \node (TA) [right = of A] {$\T{\ACat}$};
      \node (A') [below = of TA] {$\ACat$};
      \path[->] (A) edge node [above] {$\eta_\ACat$} (TA);
      \path[->] (TA) edge node [right] {$a$} (A');
      \draw[double] (A) -- (A');
    \end{tikzpicture}
    \qquad
    \DiagramSquare{
      nw = \T{\T{\ACat}},
      ne = \T{\ACat},
      se = \ACat,
      sw = \T{\ACat},
      north = \T{a},
      east = a,
      south = a,
      west = \mu_{\ACat},
      width = 2.5cm,
    }
  \end{gather*}

  The notion of morphism between $\T$-algebras differs from the 1-categorical
  case as we are primarily concerned with a weaker morphism between monad
  algebras than the 1-dimensional case.
  In particular, a morphism of monad algebras $\Mor[f]{(\ACat,a)}{(\BCat,b)}$
  is a functor $\Mor[f]{\ACat}{\BCat}$ equipped with an natural isomorphism
  $\Mor[\alpha]{b \circ \T{f}}{f \circ a}$. The natural isomorphism is required
  to satisfy certain coherence conditions, but we elide these here and refer
  the interested reader to \citet{blackwell-kelly-power:1989}.
\end{node}

\begin{node}
  The category of $\T$-algebras, weak morphisms, and natural transformations organizes into a
  2-category $\ALG{\T}$. We say that a 2-category $\ECat$ is 2-monadic over $\CAT$ when it is
  2-equivalent to the $\ALG{\T}$ for some $\T$.
\end{node}

\begin{node}
  Despite the strictness of 2-monads and their algebras, examples of 2-monads on $\CAT$ and
  $\CAT[g]$ are plentiful. For instance, the functor assigning $\CCat$ to its finite limit
  completion $\Mor|{|->}|{\CCat}{\LexCompl{\CCat}}$ is a 2-monad and the 2-category $\LEX$ is 2-monadic over
  $\CAT$. Explicitly, $\LEX$ is equivalent to the category of monad algebras for $\LexCompl{\prn{-}}$.

  Monoidal, symmetric monoidal or finitely cocomplete categories are also all 2-monadic over
  $\CAT$. The theory is developed systematically by \citet{blackwell-kelly-power:1989}, but
  generally a 2-category consisting of structured categories, structure-preserving functors, and
  natural transformations will be 2-monadic over $\CAT$.

  The exception to the rule is categories with structure like exponentials or
  dependent products, as they behave contravariantly; here the enrichment can then
  be made only over the wide subcategory $\GRPD\subseteq\CAT$.
\end{node}

\subsection{Constructing 2-monads from generators and relations}
\label{sec:sketching:finitary}

\begin{node}
  The class of \emph{finitary} 2-monads proves to be of particular importance.
  Recall that in 1-category theory, a monad is finitary when it preserves
  filtered colimits; the significance of finitary monads in a 1-category is that
  they correspond to (finitary) algebraic theories --- in other words, finitary
  1-monads can be presented by generators and relations.

  The same notions can be adapted to any presentable category, and generalized
  to a presentable 2-category (for instance, $\CAT$ and
  $\CAT[g]$)~\citep{kelly-power:1993}. This generalization allows us to
  present \emph{2-monads} by generators and relations, just like finitary
  1-monads. Many natural 2-monads are finitary, including all the examples
  presented above (finitely complete categories, monoidal categories, \emph{etc.}).
\end{node}

\begin{node}
  A more casual explanation of \emph{generators and relations} perspective on
  finitary 2-monads is given by \citet{kinoshita-power-takeyama:1999}, and we
  present some of the explanation given there but restricting to the case of
  $\CAT$ ($\CAT[g]$ is analogous).
\end{node}

\begin{node}
  We define a functor $\Mor[\Shape]{\ShapeCat}{\CAT}$ of \emph{operations},
  where $\ShapeCat$ is the discrete category of isomorphism classes of finite
  categories. The finite categories generalize the notion of \emph{arity} from
  classical algebraic theories by allowing for a finite collection of objects
  and arrows to be specified as input to an operation. Given an isomorphism
  class $\Ar{c}$ of finite categories, the category $\Shape{\Ar{c}}$ is
  defined to be the coproduct of the shape of results of all the operations with
  input arity $\Ar{c}$.

  An $\Shape$-algebra is a category $\ACat$ equipped with functors
  $\Mor[\nu\Sub{\Ar{c}}]{\Hom*{\Ar{c}}{\ACat}}{\Hom*{\Shape{\Ar{c}}}{\ACat}}$
  assigning a choice of input to the output associated with each operation.  A
  morphism of algebras $\Mor[f]{\prn{\ACat,\nu_\bullet}}{\prn{\BCat,\mu_\bullet}}$ is a functor
  between carriers that satisfies the following equation at each $\Ar{c} : \ShapeCat$:
  \[
    \DiagramSquare{
      width = 3.5cm,
      nw = \Hom*{\Ar{c}}{\ACat},
      ne = \Hom*{\Shape{\Ar{c}}}{\ACat},
      se = \Hom*{\Shape{\Ar{c}}}{\BCat},
      sw = \Hom*{\Ar{c}}{\BCat},
      north = \nu\Sub{\Ar{c}},
      east = \Hom*{\Shape{\Ar{c}}}{f},
      west = \Hom*{\Ar{c}}{f},
      south = \mu\Sub{\Ar{c}},
    }
  \]
\end{node}

\begin{node}
  In order to specify the equational part of a presentation of a finitary 2-monad, we must
  generate the derived operations of a theory; these are the operations that
  can be built by combining the various primitive operations specified by
  $\Shape$. We will define the functor of derived operations
  $\Mor[\Shape<\omega>]{\ShapeCat}{\CAT}$ by taking a direct limit over the
  derived operations of each possible finite depth.

  We define the $n$th functor of derived operations
  $\Mor[\Shape<n>]{\ShapeCat}{\CAT}$ by induction on $n\in\mathbb{N}$, writing
  $\EmbMor[J]{\ShapeCat}{\CAT}$ for the evident embedding:
  \begin{align*}
    \Shape<0>{\Ar{c}} &= J\prn{\Ar{c}}
    \\
    \Shape<n+1>{\Ar{c}} &=
    \smash{
      \underbrace{J\prn{\Ar{c}}}_{\textit{vars.}}
      +
      \underbrace{
        \overbrace{\Sum{\Ar{d} \in \ShapeCat}}^{\textit{arity}}
        \overbrace{\Shape{\Ar{d}}}^{\textit{symbol}}
        \times
        \overbrace{\Hom*{\Ar{d}}{\Shape<n>{\Ar{c}}}}^{\textit{arguments}}
      }_{\textit{operations}}
    }
  \end{align*}
  \vspace{.5em}

  There is a canonical map (often a monomorphism)
  $\Mor{\Shape<n>}{\Shape<n+1>}$; hence we may define $\Shape<\omega>$ to be
  the colimit of the resulting sequence.
  A routine computation shows that an $\Shape$-algebra structure
  $\prn{\ACat,\nu_\bullet}$ induces an $\Shape<\omega>$-algebra structure
  $\prn{\ACat,\nu^\omega_\bullet}$ by induction.
\end{node}

\begin{node}
  With $\Shape<\omega>$ in hand, we may specify the equations of a given theory
  through a functor $\Mor[\Eqns]{\ShapeCat}{\CAT}$ together with a pair of
  natural transformations $\Mor[\LHS,\RHS]{\Eqns}{\Shape<\omega>}$.
  Roughly, $\Eqns{\Ar{c}}$ specifies the shape of the equations imposed on the
  theory between terms with free variables of shape $\Ar{c}$, and the natural
  transformations $\LHS/\RHS$ interpret this shape into actual configurations
  of terms.

  An $\Shape$-algebra $\prn{\ACat, \nu_\bullet}$ satisfies the equations
  specified by $\Eqns$ when the following diagram commutes for each $\Ar{c} :
  \ShapeCat$:
  \begin{equation}
    \begin{tikzpicture}[diagram]
      \node (C) {$\Hom*{\Ar{c}}{\ACat}$};
      \node[right = 3.5cm of C] (Somega) {$\Hom*{\Shape<\omega>{\Ar{c}}}{\ACat}$};
      \node[right = 3.5cm of Somega] (E) {$\Hom*{\Eqns{\Ar{c}}}{\ACat}$};
      \path[->] (C) edge node[above] {$\nu^\omega\Sub{\Ar{c}}$} (Somega);
      \path[->] ($(Somega.east) + (0,0.1)$) edge node[above] {$\Hom*{\LHS^{\Ar{c}}}{\ACat}$} ($(E.west)+(0,0.1)$);
      \path[->] ($(Somega.east) + (0,-0.1)$) edge node[below] {$\Hom*{\RHS^{\Ar{c}}}{\ACat}$} ($(E.west)+(0,-0.1)$);
    \end{tikzpicture}
    \label[diagram]{diag:equation-sat}
  \end{equation}

  Specializing again to classical algebraic theories, this diagram commutes
  when $\ACat$ satisfies the equations specified by $\Eqns$ after instantiating
  the free variables with arbitrary elements of $\ACat$.
\end{node}

\begin{node}
  If $\prn{\ACat,\nu}$  is a $\Shape$-algebra for which
  \cref{diag:equation-sat} commutes for each $\Ar{c}$, we call it an
  $\gls{\Shape,\Eqns}$-algebra. We take the category of
  $\gls{\Shape,\Eqns}$-algebras to be the evident full subcategory of
  $\Shape$-algebras.
\end{node}

\begin{node}
  For any $\gls{\Shape,\Eqns}$, there exists a finitary 2-monad $\T$ and an equivalence between the
  category of $\gls{\Shape,\Eqns}$-algebras and $\T$-algebras.
\end{node}

\begin{node}\label{node:lcccs-are-monadic}
  The machinery exposed in this section directly implies that locally Cartesian
  closed categories can be characterized by a finitary 2-monad on $\CAT[g]$.
  By a general result, therefore, $\LCCC{}$ is bi-cocomplete and
  pseudo-complete~\citep{kelly:1989,blackwell-kelly-power:1989}.\footnote{In
  fact, $\LCCC{}$ is closed under PIE-limits~\citep{nlab:pie-limit}.}
\end{node}

\subsubsection{An extended example: \texorpdfstring{$\LEX$}{LEX}}
\label{sec:sketching:lex}

\CounterZeroNext{node}

\begin{node}
  For expository purposes we demonstrate how to construct the 2-monad presenting
  $\LEX$, the 2-category of finitely complete categories.
\end{node}

\begin{node}
  It is well-known that a category is finitely complete when it contains a
  terminal object and all pullbacks. Both of these conditions can be encoded as
  a pair of operations, with one operation producing the terminal object (resp.
  pullback) and a second equipping it with the distinguished maps provided by
  its universal property.
\end{node}

\begin{node}\label{node:terminal-object}
  \emph{The terminal object.}
  The operation that produces the terminal object
  requires no input and produces a single object as output. The universal property takes a single
  object $X$ and extends it to a map $\Mor{X}{\ObjTerm{}}$. Both of these are encoded as operations:
  \[
    \Shape{\brc{}} = \brc{\CodeObjTerm}
    \qquad
    \Shape{\brc{\bullet}} = \brc{\Mor[\CodeArrTerm{\bullet}]{\Dom{\CodeArrTerm{\bullet}}}{\Cod{\CodeArrTerm{\bullet}}}}
  \]
\end{node}

\begin{node}\label{node:pullbacks}
  \emph{Pullbacks.}
  We require a similar set of operations for pullbacks, though they take a span of objects as input,
  and produce a commuting diagram as output:
  \[
    \Shape{\brc{\CoSpan{\bullet}{\bullet}{\bullet}}} = \CCat
    \qquad
    \Shape{\Ar{d}} = \DCat
  \]

  We define $\CCat$, $\Ar{d}$ and $\DCat$ to be the following categories:
  \[
    \CCat = \Ar{d} =
    \begin{tikzpicture}[diagram, node distance = 1cm, baseline = -0.55cm]
      \SpliceDiagramSquare{
        ne = \bullet,
        nw = \bullet,
        se = \bullet,
        sw = \bullet,
        width = 1cm,
        height = 1cm,
      }
    \end{tikzpicture}
    \qquad
    \DCat =
    \begin{tikzpicture}[diagram, node distance = 1cm]
      \SpliceDiagramSquare{
        ne = \bullet,
        nw = \bullet,
        se = \bullet,
        sw = \bullet,
        width = 1cm,
        height = 1cm,
      }
      \node[above left = 0.75cm and 0.75cm of nw] (x) {$\bullet$};
      \path[->, bend right] (x) edge (sw);
      \path[->, bend left] (x) edge (ne);
      \path[->] (x) edge (nw);
    \end{tikzpicture}
  \]
\end{node}

\begin{node}
  On their own, the operations specified in
  \cref{node:terminal-object,node:pullbacks} are not sufficient to ensure that
  an algebra has terminal objects and pullbacks.
  To be concrete, let $\prn{\ACat,\nu}$ be an algebra for $\Shape$. There is a
  distinguished object $\CodeObjTerm$ induced by the first operation of
  \cref{node:terminal-object} and a choice of map
  $\Mor[\CodeArrTerm{A}]{\Dom{\CodeArrTerm{A}}}{\Cod{\CodeArrTerm{A}}}$ for each $A : \ACat$ induced by the second
  operation; so far, nothing we have done has ensured that the fixed boundary
  of $\CodeArrTerm{A}$ to be anything in particular.

  Recall that in the enriched setting, the algebra maps
  $\Mor[\nu\Sub{\Ar{c}}]{\Hom*{\Ar{c}}{\ACat}}{\Hom*{\Shape{\Ar{c}}}{\ACat}}$
  making $\ACat$ an algebra are functors, not merely functions between sets.
  The functoriality conditions on maps ensure that certain naturality conditions between operators
  are automatically enforced. In the case of the first operation the condition is trivial, but for
  the second operation we obtain a commuting square for each map $\Mor[f]{A}{B}$:

  \begin{equation}\label[diagram]{eq:arr-term-functorial}
    \DiagramSquare{
      nw = \Dom{\CodeArrTerm{A}},
      ne = \Dom{\CodeArrTerm{B}},
      sw = \Cod{\CodeArrTerm{A}},
      se = \Cod{\CodeArrTerm{B}},
      north = \CodeArrTerm{f}^0,
      south = \CodeArrTerm{f}^1,
      west = \CodeArrTerm{A},
      east = \CodeArrTerm{B},
    }
  \end{equation}

  To make $\CodeObjTerm$ into a terminal object is, then, we must add \emph{equations} that ensure
  $\Dom{\CodeArrTerm{A}} = A$ and $\Cod{\CodeArrTerm{A}} = \CodeObjTerm$, and to ensure that
  $\CodeArrTerm{A}$ is the unique map with this property. We explore this process in
  \cref{node:trm-obj:eqns:bdry,node:trm-obj:eqns:uniq}.
\end{node}

\begin{node}\label{node:trm-obj:eqns:bdry}
  We will add equations to ensure that $\CodeArrTerm{A}$ has the correct boundary. In fact, we
  will do more than this and ensure that $\CodeArrTerm{}$ has the correct functorial behavior by
  adding two equations: one ensuring for every $\Mor[f]{A}{B}$ that $\CodeArrTerm{f}^0 = f$ and the
  other ensures $\CodeArrTerm{f}^1 = \ArrId{\CodeObjTerm}$.
  We prepare to add these two equations by defining $\Eqns$ at
  $\brc{\Mor{\bullet}{\bullet}}$ to be the following category:
  \[
    \Eqns(\brc{\Mor{\bullet}{\bullet}}) =
    \brc{
      \begin{tikzpicture}[inline diagram]
        \node (A) {$\bullet$};
        \node [right= 2.5cm of A] (B) {$\bullet$};
        \path[->] (A) edge node[above] {$\Code{t/eq/f-dom}$} (B);
      \end{tikzpicture}
      \qquad
      \begin{tikzpicture}[inline diagram]
        \node (A) {$\bullet$};
        \node [right= 2.5cm of A] (B) {$\bullet$};
        \path[->] (A) edge node[above] {$\Code{t/eq/f-cod}$} (B);
      \end{tikzpicture}
    }
  \]

  Recall that $\LHS,\RHS$ must be natural transformations from $\Eqns$ to $\Shape<\omega>$; this
  means that the component of $\LHS$ at $\brc{\Mor{\bullet}{\bullet}}$ must send an equation symbol
  from $\Eqns{\brc{\Mor{\bullet}{\bullet}}}$ to a derived term with one free variable. Hence, we
  extend the definition of $\LHS,\RHS$ at $\brc{\Mor{\bullet}{\bullet}}$ to specify the two
  equations:
  \begin{gather*}
    \begin{aligned}
      \LHS{\Code{t/eq/f-dom}} &=
      \prn{\Mor[f]{A}{B} \vdash \CodeArrTerm{f}^0}\\
      \RHS{\Code{t/eq/f-dom}} &=
      \prn{\Mor[f]{A}{B} \vdash f}
    \end{aligned}
    \qquad
    \begin{aligned}
      \LHS{\Code{t/eq/f-cod}} &=
      \prn{\Mor[f]{A}{B} \vdash \CodeArrTerm{f}^1}\\
      \RHS{\Code{t/eq/f-cod}} &=
      \prn{\Mor[f]{A}{B} \vdash \ArrId{\CodeObjTerm}}
    \end{aligned}
  \end{gather*}
\end{node}

\begin{node}\label{node:trm-obj:eqns:uniq}
  We also wish to ensure that $\CodeArrTerm{A}$ is the unique arrow from
  $\Mor{A}{\CodeObjTerm}$. This is done in a somewhat roundabout manner: we add an equation with no
  free variables ensuring that $\Mor[\CodeArrTerm{\CodeObjTerm{}}]{\CodeObjTerm{}}{\CodeObjTerm{}}$
  is the identity.
  When \cref{eq:arr-term-functorial} is instantiated with $\Mor[f]{A}{\CodeObjTerm}$, we have
  $\CodeArrTerm{\CodeObjTerm} \circ \CodeArrTerm{f}^0 = \CodeArrTerm{f}^1 \circ \CodeArrTerm{A}$.
  We have just required $\CodeArrTerm{\CodeObjTerm} = \ArrId{\CodeObjTerm}$ and
  \cref{node:trm-obj:eqns:bdry} ensures $\CodeArrTerm{f}^0 = f$ and
  $\CodeArrTerm{f}^1 = \ArrId{\CodeObjTerm}$. Thus, we have $f = \CodeArrTerm{A}$ as required.
\end{node}

\begin{node}
  The equations for the pullback operations are similar. For each cospan
  $\CoSpan{A_2}{A_0}{A_1}$ we obtain a square, and for each square we obtain a
  ``gap map''; we add equations to fix the boundary of this square to extend
  the cospan, and to fix the boundary of the gap map. Finally, an equation is added to
  force the gap map of the purported pullback square itself to be the identity;
  combined with naturality this equation ensures that the gap map is unique, and hence that the
  square is in fact a pullback square.
\end{node}

\begin{node}
  In the case of (locally) Cartesian closed categories, the enrichment is taken
  over $\CAT[g]$, not $\CAT$; this is a consequence of the contravariance
  involved in exponentials. Accordingly the functorial conditions for the
  operations are far weaker, with naturality squares guaranteed only for
  isomorphisms.

  The consequence is that \cref{eq:arr-term-functorial} must be added as an
  additional equation when defining (local) cartesian closure as a 2-monad,
  because it would no longer directly follow from the enrichment. We would also
  similarly add equations to enforce the appropriate functoriality of the
  transposition operation for (local) exponentials.
\end{node}

\section{Sketching theories in a doctrine}
\label{sec:sketching}

\begin{node}
  In \cref{sec:sketching:finitary} we outlined the proof that $\LCCC$ is
  finitarily 2-monadic over $\CAT[g]$; hence we may define the free locally
  Cartesian closed category $\T{\CCat}$ on any category $\CCat$.  As a free
  object, $\T{\CCat}$ enjoys the following universal property: naturally in
  a given locally Cartesian closed category $\ECat$, we have
  $\Hom[\CAT[g]]{\CCat}{\ECat} \cong \Hom[\LCCC]{\T{\CCat}}{\ECat}$. In
  other words, locally Cartesian closed functors out of $\T{\CCat}$ are
  completely determined by their restriction to $\CCat$.
\end{node}

\begin{node}\label{ex:walking-tt}
  On its own, the 2-monad of local Cartesian closure already allows us to
  construct particularly simple type theories.  For instance, the \emph{walking
  type theory} generated by a single morphism $\Mor[\El]{\EL}{\TY}$ can be
  defined by the free locally Cartesian closed category generated by a single
  morphism.
\end{node}

\begin{node}
  \label{node:sketches-are-necessary}
  In order to construct more realistic theories, however, we must be able to
  specify particular connectives whose boundaries involve dependent products
  and pullbacks. For instance, to specify the closure of the ``walking type
  theory'' \cref{ex:walking-tt} under dependent products, one needs to add a
  cartesian morphism $\Mor{P_\El\prn{\El}}{\El}$ where $P_\El$ is the
  polynomial endofunctor on $\El$ --- but the definition of $P_\El$ mentions dependent
  products:
  \[
    P_{\El}\prn{X} = \Sum{A:\TY}{\Prod{x : \El\brk{A}}{X}}
  \]

  In essence the difficulty is that one seems to need to interleave the
  specification of generators with the free extension by locally Cartesian
  closed structure. One way to do this is using a syntactic logical framework;
  an alternative method, which we explain here, is the theory of
  \emph{sketches}~\citep{lawvere:thesis,ehresmann:1966,ehresmann:structured-categories:1968,ehresmann:sketches:1968}.\footnote{In
  some traditional accounts, the data of a sketch is concentrated in a
  \emph{directed graph} equipped with a choice various subgraphs that will be
  realized by commutative diagrams in a category. We follow
  \citet{kinoshita-power-takeyama:1999} in starting directly from categories
  and ignoring the directed graph structure; our choices reflects the fact that
  it is not so difficult to freely generate a category by a collection of
  diagrams.}
\end{node}

\begin{node}
  The classical theory of sketches is restricted to developing theories
  involving finite (co)limits, but \citet{kinoshita-power-takeyama:1999} have
  given a generalization that works for 2-monads more generally. We now
  briefly recall the classical theory of finite limit sketches before
  introducing this generalization, and refer the reader to
  \citet{adamek-rosicky:1994} for a textbook account.
\end{node}

\subsection{Finite limit sketches}

\NewDocumentCommand{\Cone}{m}{#1\Sup{\lhd}}

\begin{node}
  Let $\DCat$ be a category; the we write $\Cone{\DCat}$ for the \emph{cone}
  above $\DCat$, which is the \emph{join} of categories $
  \brc{\ObjInit}\mathbin{\star}\DCat$ in the sense of \citet{lurie:2009}.
  Explicitly $\Cone{\DCat}$ freely adjoins an initial object to $\DCat$.
\end{node}

\begin{node}
  Given a category $\CCat$, a \emph{finite cone} in $\CCat$ is just a
  functor $\Mor{\Cone{\Ar{d}}}{\CCat}$ for some finitary category $\Ar{d}$.
\end{node}

\begin{node}\label{node:finite-limit-sketch}
  A \emph{finite limit sketch} is defined to be a category $\SketchCat$ (not
  necessarily finitely complete) together with a collection of finite cones
  $\brc{\Mor[d_i]{\Cone{\Ar{d}_i}}{\SketchCat}}$; the cones $d_i$ are called
  \emph{marked}. We will often use $\SketchCat$ metonymically to refer to the
  pair $\prn{\SketchCat,d_\bullet}$.
\end{node}

\begin{node}
  The purpose of $\SketchCat$ is to specify the generating objects and
  morphisms for our theory; because the (intended) boundaries of generating
  morphisms frequently involve not only generating objects but finite limits
  thereof, however, we will include ``dummy generators'' for these boundaries
  that will subsequently be \emph{marked} in the sense of
  \cref{node:finite-limit-sketch}. This is the solution that the language of
  sketches offers to the problem of interleaving specification with free
  generation.

  A finite cone that is ``marked'' does not necessarily have a universal
  property in $\SketchCat$, so there are no conditions to check when marking a
  cone. The chosen markings are then put to use in the definition of
  \emph{models} of a sketch in \cref{node:model-of-lex-sketch}, in which marked cones are required to be realized
  by actual limit cones.
\end{node}

\begin{node}\label{node:model-of-lex-sketch}
  We define a model of $\SketchCat$ in a finitely complete category $\ECat$ to
  be a functor $\Mor[f]{\SketchCat}{\ECat}$ such that $f \circ d_i$ is limiting
  for each marked cone $d_i$.  We write $\MOD{\SketchCat{}}{\ECat}$ for the
  category of $\ECat$-valued models of $\SketchCat$ and natural transformations between them.
\end{node}

\begin{node}
  In practice, this means that while the sketch of a group might not contain the product
  $G \times G$, we can mark the domain of the multiplication operation $\Mor[m]{P}{G}$ so that any
  model of the sketch must realize $P$ as $G \times G$.
\end{node}

\begin{node}
  \label{node:lex-sketch-cl-cat}
  The central theorem of finite limit sketches states that there exists a finite limit category
  that completes a sketch $\SketchCat$ by adding the missing limits while ensuring that marked
  cones become proper limits. More precisely, there is a functor from the category of sketches to
  $\LEX$ that induces the following equivalence of categories:
  \[
    \Hom[\LEX]{\Thy{\SketchCat}}{\CCat}
    \simeq
    \MOD{\SketchCat}{\CCat}
  \]
\end{node}

\subsection{Generalized sketches}

\begin{node}
  In order to generalize \cref{node:lex-sketch-cl-cat} to locally Cartesian closed categories, we
  will replace finite limits with an arbitrary finitary $2$-monad $\T$ following the work of
  \citet{kinoshita-power-takeyama:1999}. In particular, we generalize the theory of sketches to
  allow for diagrams to be marked with structure from $\T$, and replace finitely complete categories
  with $\T$-algebras.

  When we instantiate $\T$ with the presentation of $\LEX$ discussed in \cref{sec:sketching:lex}, the
  results in the more general formulation will specialize to those of \cref{node:lex-sketch-cl-cat}.
\end{node}

\begin{node}
  The most complex piece of the generalized sketching machinery is the replacement for marking
  limits. We accomplish this by breaking a marked diagram into two components. Rather than a single
  finitary category with an initial object, we require a pair of finitary categories
  $\prn{\Ar{c},\Ar{d}}$ along with a commuting triangle:
  \begin{equation}
    \begin{tikzpicture}[diagram,baseline = (Tc.base)]
      \node (d) {$\Ar{d}$};
      \node [right = 3cm of d] (Tc) {$\T{\Ar{c}}$};
      \node [above = 1.5cm of d] (c) {$\Ar{c}$};
      \path[->] (d) edge node[below] {$k$} (Tc);
      \path[->] (c) edge node[sloped, above] {$\eta_{\Ar{c}}$} (Tc);
      \path[->] (c) edge node[left] {$j$} (d);
    \end{tikzpicture}
    \label[diagram]{diag:generalized-marked-diagram}
  \end{equation}

  The finitary category $\Ar{c}$ contains the \emph{generators} involved in the
  marked diagram; $\Ar{d}$ is the actual shape of the diagram, and the functor
  $\Mor[k]{\Ar{d}}{\T{\Ar{c}}}$ explains how the diagram is intended to be
  realized in terms of the $\T$-structure generated by $\Ar{c}$.  The fact that
  \cref{diag:generalized-marked-diagram} commutes is a well-formedness
  condition on a marked diagram to ensure that the ``generators'' are taken to
  their avatars in the realization.
\end{node}

\begin{node}
  A sketch for a 2-monad $\T$ is a category $\SketchCat$ equipped with a set of quadruples
  $\Compr{(\Ar{c}_i,\Ar{d}_i,j_i,k_i)}{k_i \circ j_i = \eta_{\Ar{c}_i}}$ in the sense of \cref{diag:generalized-marked-diagram}
  together with functors functors $\Mor[\phi_i]{\Ar{d}_i}{\SketchCat}$.
\end{node}

\begin{node}\label{node:sketch-model}
  A model of a sketch $\SketchCat$ in a $\T$-algebra $\prn{\ACat, a}$ is a functor $\Mor[f]{\SketchCat}{\ACat}$
  together with a collection of natural isomorphisms $\alpha_i$ witnessing the weak commutativity of
  the black square in \cref{diag:sketch-model:0} below:
  \begin{equation}\label[diagram]{diag:sketch-model:0}
    \begin{tikzpicture}[diagram,baseline=(sq/sw.base)]
      \SpliceDiagramSquare<sq/>{
        width = 5cm,
        north/node/style = upright desc,
        height = 1.5cm,
        nw = \Ar{d}_i,
        sw = \SketchCat,
        ne = \T{\ACat},
        se = \ACat,
        north = \text{\small$\T{f \phi_i j_i} \circ k_i$},
        south = f,
        east = a,
        west = \phi_i
      }
      \node [between = sq/nw and sq/se] {$\alpha_i$};
      \node (nw) [gray, above = 1.5cm of sq/nw] {$\T{\Ar{c}_i}$};
      \node (ne) [gray, above = 1.5cm of sq/ne] {$\T{\SketchCat}$};
      \node (n) [gray, between = nw and ne] {$\T{\Ar{d}_i}$};
      \path[->,gray,] (sq/nw) edge node [left] {$k_i$} (nw);
      \path[->,gray,] (nw) edge node [above] {$\T{j_i}$} (n);
      \path[->,gray,] (n) edge node [above] {$\T{\phi_i}$} (ne);
      \path[->,gray,] (ne) edge node [right] {$\T{f}$} (sq/ne);
    \end{tikzpicture}
  \end{equation}

  The diagram above states, in essence, that the marked diagrams in
  $\SketchCat$ are taken by the functor $f$ to the \emph{actual} structures
  they are intended to denote.
  We additionally require that filler $\alpha_i$
  restricts to the identity along $\Mor[j_i]{\Ar{c}_i}{\Ar{d}_i}$; in particular,
  the black square in \cref{diag:sketch-model:1} below commutes on the
  nose:
  \begin{equation}\label[diagram]{diag:sketch-model:1}
    \begin{tikzpicture}[diagram,baseline=(sq/sw.base)]
      \SpliceDiagramSquare<sq/>{
        width = 5cm,
        north/node/style = upright desc,
        height = 1.5cm,
        nw = \Ar{d}_i,
        sw = \SketchCat,
        ne = \T{\ACat},
        se = \ACat,
        north = \text{\small$\T{f \phi_i j_i} \circ k_i$},
        south = f,
        east = a,
        west = \phi_i,
        west/style = gray,
        west/node/style = upright desc,
      }
      \node [between = sq/nw and sq/se] {$\alpha_i$};
      \node (nw) [gray, above = 1.5cm of sq/nw] {$\T{\Ar{c}_i}$};
      \node (ne) [gray, above = 1.5cm of sq/ne] {$\T{\SketchCat}$};
      \node (n) [gray, between = nw and ne] {$\T{\Ar{d}_i}$};
      \node [gray,between = sq/nw and sq/se] {$\alpha_i$};
      \node (nww) [left = of sq/nw] {$\Ar{c}_i$};
      \path[->] (nww) edge node [upright desc] {$j_i$} (sq/nw);
      \path[->] (nww) edge node [sloped,below] {$\phi_i\circ j_i$} (sq/sw);
      \path[->,gray] (nww) edge node [sloped,above] {$\eta\Sub{\Ar{c}_i}$} (nw);
      \path[->,gray] (ne) edge node [right] {$\T{f}$} (sq/ne);

      \path[->,gray,] (sq/nw) edge node [upright desc] {$k_i$} (nw);
      \path[->,gray,] (nw) edge node [above] {$\T{j_i}$} (n);
      \path[->,gray,] (n) edge node [above] {$\T{\phi_i}$} (ne);
      \path[->,gray,] (ne) edge node [right] {$\T{f}$} (sq/ne);
    \end{tikzpicture}
  \end{equation}
\end{node}

\begin{node}
  The purpose of the second condition of \cref{node:sketch-model} governing the
  2-cells $\alpha_i$ can be understood in plain English.  The finite category
  $\Ar{d}_i$ specifies the shape of a ``derived arity'' that will be realized
  by certain $\T$-algebra structure; the 2-cell $\alpha_i$ allows the
  realization of this arity to be off by an isomorphism, but
  the additional condition that \cref{diag:sketch-model:1} commute on the nose
  ensures that the \emph{generators} involved in this derived arity are realized in
  $\ACat$ exactly by their interpretation under $f$.

  The balance of strictness and weakness embodied in \cref{node:sketch-model}
  is very important in practice: it would be absurd to insist that
  \emph{(e.g.)} some marked cone be realized by the exact choice of limits
  determined under the algebra $\Mor{\T{\ACat}}{\ACat}$. But when this marked
  cone involves a generator, it is perfectly sensible to insist that the
  realization involve this \emph{exact} generator, and not some isomorph of it.
\end{node}

\begin{node}
  Models of a sketch in a particular $\T$-algebra assemble into a category that
  we will call $\MOD{\SketchCat}{\ACat}$ whose morphisms are given by natural
  transformations in $\ACat$ that commute appropriately commute with the generators
  of the sketch. We refer the reader to \citet[Definition
  4.4]{kinoshita-power-takeyama:1999} for details.
\end{node}

\begin{node}\label{node:generalized-sketch}
  The generalization of \cref{node:lex-sketch-cl-cat} now states that there
  exists a functor $\Thy$ which assigns to each sketch $\SketchCat$ a
  $\T$-algebra satisfying the following equivalence:
  \[
    \Hom[\ALG{\T}]{\Thy{\SketchCat}}{A}
    \simeq
    \MOD{\SketchCat}{A}
  \]
\end{node}

\begin{node}\label{node:sketchable}
  In particular, the machinery of sketches applies to the finitary 2-monad for
  locally Cartesian closed categories. While the fully formal definition of a
  sketch of any kind is quite technical, at an informal level the theory of sketches
  allows one to freely use \emph{(e.g.)} dependent products and finite limits
  while describing the generators of a free locally Cartesian closed category.
\end{node}

\section{Sketching type theories}
\label{sec:sketching:examples}

\begin{node}
  We now demonstrate how the syntactic categories (categories of judgments) of
  a variety of type theories may be quickly sketched using this machinery.
\end{node}

\begin{node}
  \label{node:el-sketch}
  First, as was mentioned in the previous section the sketch of the ``walking type theory''
  $\SketchCat$ can be presented by a category with a single generating morphism
  $\Mor[\El]{\EL}{\TY}$. The universal property of $\Thy{\SketchCat}$ states that a locally Cartesian closed morphism
  $\Mor{\SketchCat}{\ECat}$ is determined precisely by a morphism $\Mor{E}{B}$ in $\ECat$.
\end{node}

\subsection{Sketching dependent products}\label{sec:pi-sketch}

\begin{node}
   We now show how one may sketch closure under dependent products. First, we extend $\SketchCat$ to
   the category generated by the following square:
   \begin{equation}\label[diagram]{diag:pi-generators}
     \DiagramSquare{
       nw = \EL^\Pi,
       sw = \TY^\Pi,
       ne = \EL,
       se = \TY,
       east = \El,
       west = \El^\Pi,
       south = \mathsf{pi},
       north = \mathsf{lam},
     }
   \end{equation}

   We must carry out \emph{two} separate markings, first to ensure that
   \cref{diag:pi-generators} is cartesian, and second to cause the left-hand
   map to be realized by the ``generic dependent product family''.
\end{node}

\begin{node}
  First we will mark \cref{diag:pi-generators} as a pullback square. The generating shape
  $\Ar{c}$ is the walking cospan, and the diagram shape $\Ar{d}$ is the walking
  square, and the functor $\Mor[j]{\Ar{c}}{\Ar{d}}$ immerses one into the other
  like so:
  \[
    \DiagramSquare{
      nw = \cdot,
      ne = j\prn{\Code{ne}},
      se = j\prn{\Code{se}},
      sw = j\prn{\Code{sw}},
      nw = \Code{nw},
      east = j\prn{\Code{e}},
      south = j\prn{\Code{s}},
      north = \Code{n},
      west = \Code{w},
      ne/style = {color = RedDevil},
      se/style = {color = RedDevil},
      sw/style = {color = RedDevil},
      east/style = {color = RedDevil},
      south/style = {color = RedDevil},
    }
  \]

  The diagram $\Mor[k]{\Ar{d}}{\T{\Ar{c}}}$ is then the following formal
  pullback square:
  \[
    \DiagramSquare{
      ne = \eta_{\Ar{c}}\prn{\Code{ne}},
      se = \eta_{\Ar{c}}\prn{\Code{se}},
      sw = \eta_{\Ar{c}}\prn{\Code{sw}},
      nw = k\prn{\Code{nw}},
      south = \eta_{\Ar{c}}\prn{\Code{s}},
      east = \eta_{\Ar{c}}\prn{\Code{e}},
      west = k\prn{\Code{w}},
      north = k\prn{\Code{n}},
      width = 2.25cm,
      north/style = exists,
      west/style = exists,
      nw/style = pullback,
      ne/style = {color = RedDevil},
      se/style = {color = RedDevil},
      sw/style = {color = RedDevil},
      east/style = {color = RedDevil},
      south/style = {color = RedDevil},
    }
  \]

  Finally, the diagram $\Mor[\phi]{\Ar{d}}{\SketchCat}$ is \cref{diag:pi-generators} itself.
\end{node}

\begin{node}
  Next we mark the left-hand map of \cref{diag:pi-generators} to be realized in
  models by $P_\El\prn{\El}$ where $P_\El$ is the polynomial endofunctor of
  $\El$ as explained by \citet{awodey:2018:natural-models}. In this case the
  generating shape $\Ar{c}$ is the walking arrow
  $\brc{\Mor[\Code{f}]{\Code{s}}{\Code{t}}}$, the diagram shape $\Ar{d}$ is a
  (disjoint) pair of arrows, and the functor $\Mor[j]{\Ar{c}}{\Ar{d}}$
  identifies $\Ar{c}$ with the right-hand arrow in $\Ar{d}$:
  \[
    \begin{tikzpicture}[diagram]
      \node (nw) {$\Code{u}$};
      \node (sw) [below = 1.5cm of nw] {$\Code{v}$};
      \node (ne) [color = RedDevil, right = 1.5cm of nw] {$j\prn{\Code{s}}$};
      \node (se) [color = RedDevil, below = 1.5cm of ne] {$j\prn{\Code{t}}$};
      \path[->] (nw) edge node [left] {$\Code{g}$} (sw);
      \path[->,color = RedDevil] (ne) edge node [right] {$j\prn{\Code{f}}$} (se);
    \end{tikzpicture}
  \]

  The diagram $\Mor[k]{\Ar{d}}{\T{\Ar{c}}}$ is then the following pair of maps in $\T{\Ar{c}}$:
  \[
    \begin{tikzpicture}[diagram]
      \node (nw) {$P_{\eta_{\Ar{c}}\prn{\Code{f}}}\prn{\eta_{\Ar{c}}\prn{\Code{s}}}$};
      \node (sw) [below = 1.5cm of nw] {$P_{\eta_{\Ar{c}}\prn{\Code{f}}}\prn{\eta_{\Ar{c}}\prn{\Code{t}}}$};
      \node (ne) [color = RedDevil, right = 2.5cm of nw] {$\eta_{\Ar{c}}\prn{\Code{s}}$};
      \node (se) [color = RedDevil, below = 1.5cm of ne] {$\eta_{\Ar{c}}\prn{\Code{t}}$};
      \path[->,exists] (nw) edge node [left] {$P_{\eta_{\Ar{c}}\prn{\Code{f}}}\prn{\eta_{\Ar{c}}\prn{\Code{f}}}$} (sw);
      \path[->,color = RedDevil] (ne) edge node [right] {$\eta_{\Ar{c}}\prn{\Code{f}}$} (se);
    \end{tikzpicture}
  \]

  Finally the diagram $\Mor[\phi]{\Ar{d}}{\SketchCat}$ is the following pair of maps:
  \[
    \begin{tikzpicture}[diagram]
      \node (nw) {$\EL^\Pi$};
      \node (sw) [below = 1.5cm of nw] {$\TY^\Pi$};
      \node (ne) [right = 1.5cm of nw] {$\EL$};
      \node (se) [below = 1.5cm of ne] {$\TY$};
      \path[->] (nw) edge node [left] {$\El^\Pi$} (sw);
      \path[->] (ne) edge node [right] {$\El$} (se);
    \end{tikzpicture}
  \]
\end{node}

\begin{node}
  The universal property of this extended sketch now ensures that a locally Cartesian closed functor
  $\Mor{\Thy{\SketchCat}}{\ECat}$ is precisely equivalent to a realization of \cref{diag:pi-generators}
  as an actual dependent product classification situation in $\ECat$.
\end{node}

\subsection{Sketching non-universal connectives}

\begin{node}
  Both dependent sums and products are governed by a universal property, which
  enables terse encodings in both natural models~\citep{awodey:2018} and the
  framework exposed here. Types not determined by a universal property, such as
  the (weak) booleans, are more challenging to encode and cannot be captured by a
  single cartesian square.
  We begin by adding the formation and introduction rules for booleans to $\SketchCat$.
  \begin{equation}\label[diagram]{diag:bool-generators:0}
    \begin{tikzpicture}[diagram,baseline=(se.base)]
      \node (nw) {$\ObjTerm$};
      \node (ne) [right = of nw] {$\EL$};
      \node (se) [below = of ne] {$\TY$};
      \path[->] (ne) edge node [right] {$\El$} (se);
      \path[->] (nw) edge node [above] {$\TmTrue,\TmFalse$} (ne);
      \path[->] (nw) edge node [sloped,below] {$\TyBool$} (se);
    \end{tikzpicture}
  \end{equation}
\end{node}

\begin{node}
  First we mark $\ObjTerm{}$ to ensure that it is realized by the terminal
  object; the generating shape $\Ar{c}$ is the empty category, and the marking
  shape $\Ar{d}$ is the terminal category $\brc{\bullet}$. We set
  $k\prn{\bullet} = \ObjTerm{\T{\Ar{c}}}$, and set $\phi\prn{\bullet} =
  \ObjTerm$.
\end{node}

\begin{node}
  In what follows, we will write $\IHom{X}{Y}$ for the exponential $Y^X$ when it exists.
\end{node}

\begin{node}\label{node:bool-elim-intuition}
  The purpose of the elimination rule for the booleans is to provide a way to
  construct elements $f : \IHom{\El\brk{\TyBool}}{\EL}$; because we are not considering
  a \emph{strict} universal property for the booleans, we will characterize
  this object only weakly. The elimination rule for the booleans states that
  there exists a section to the ``cartesian gap map'' of the following
  square:
  \begin{equation}\label[diagram]{diag:bool-elim:0}
    \DiagramSquare{
      width = 3.5cm,
      nw = \IHom{\El\brk{\TyBool}}{\EL},
      ne = \EL\times\EL,
      se = \TY\times\TY,
      east = \El\times\El,
      north = \prn{\TmTrue^*, \TmFalse^*},
      south = \prn{\TmTrue^*, \TmFalse^*},
      west = \IHom{\El\brk{\TyBool}}{\El},
      sw = \IHom{\El\brk{\TyBool}}{\TY},
    }
  \end{equation}

  Unfolding into more elementary terms, the eliminator would be the section $s$ below:
  \begin{equation}\label[diagram]{diag:bool-elim:1}
    \begin{tikzpicture}[diagram,baseline=(sq/sw.base)]
      \SpliceDiagramSquare<sq/>{
        width = 3.5cm,
        nw/style = pullback,
        nw = \mathsf{Elim},
        ne = \EL\times\EL,
        se = \TY\times\TY,
        east = \El\times\El,
        south = \prn{\TmTrue^*, \TmFalse^*},
        north = q,
        west = p,
        sw = \IHom{\El\brk{\TyBool}}{\TY},
        west/node/style = upright desc,
        north/node/style = upright desc,
      }
      \node (nw) [above left = 2.5cm of sq/nw] {$\IHom{\El\brk{\TyBool}}{\EL}$};
      \path[->,bend right=30] (nw) edge node [left] {$\IHom{\El\brk{\TyBool}}{\El}$} (sq/sw);
      \path[->,bend left=30] (nw) edge node [sloped,above] {$\prn{\TmTrue^*,\TmFalse^*}$} (sq/ne);
      \path[->,bend left=20] (nw) edge node [desc] {$r$} (sq/nw);
      \path[->,bend left=20,exists,] (sq/nw) edge node [desc] {$s$} (nw);
    \end{tikzpicture}
  \end{equation}
\end{node}

\begin{node}\label{node:sketch-bool-elim}
  We may sketch the situation described in \cref{node:bool-elim-intuition},
  because it involves only pullbacks and exponentials. It will be simpler if we
  first eliminate the combination of product and pullback from \cref{diag:bool-elim:1} in terms of a single
  finite limit; we first add the following objects and morphisms to
  $\SketchCat$ such that $s$ is a section of $r$:
  \begin{equation}\label[diagram]{diag:bool-generators:1}
    \begin{tikzpicture}[diagram,baseline=(U-l.base)]
      \node (Elim) {$\mathsf{Elim}$};
      \node (U/bool) [below = of Elim] {$\Sl{\TY}{\TyBool}$};
      \node (E-r) [right = of Elim] {$\EL$};
      \node (E-l) [left = of Elim] {$\EL$};
      \node (U-r) [right = of U/bool] {$\TY$};
      \node (U-l) [left = of U/bool] {$\TY$};
      \node (E/bool) [above = of Elim] {$\Sl{\EL}{\TyBool}$};
      \path[->] (Elim) edge node [upright desc] {$p$} (U/bool);
      \path[->] (E-l) edge node [left] {$\El$} (U-l);
      \path[->] (E-r) edge node [right] {$\El$} (U-r);
      \path[->] (U/bool) edge node [below] {$h_0$} (U-l);
      \path[->] (U/bool) edge node [below] {$h_1$} (U-r);
      \path[->] (Elim) edge node [upright desc] {$q_0$} (E-l);
      \path[->] (Elim) edge node [upright desc] {$q_1$} (E-r);
      \path[->] (E/bool) edge node [sloped,above] {$\dot{h}_0$} (E-l);
      \path[->] (E/bool) edge node [sloped,above] {$\dot{h}_1$} (E-r);
      \path[->,bend left=20] (E/bool) edge node [upright desc] {$r$} (Elim);
      \path[->,bend left=20] (Elim) edge node [upright desc] {$s$} (E/bool);
    \end{tikzpicture}
  \end{equation}

  Next we will mark \cref{diag:bool-generators:1}, setting the generating shape
  $\Ar{c}$ to be the shape of \cref{diag:bool-generators:0} and setting the
  marking shape $\Ar{d}$ to be the disjoint union of $\Ar{c}$ with the shape of
  \cref{diag:bool-generators:1} omitting $s$. The diagram
  $\Mor[j]{\Ar{c}}{\Ar{d}}$ is obvious, so we focus on defining
  $\Mor[k]{\Ar{d}}{\T{\Ar{c}}}$; the restriction of $k$ along $j$ is already
  fixed, so it remains to choose the following subdiagram, writing $\floors{-}$
  for $\eta_{\Ar{c}}$:
  \begin{equation}\label[diagram]{diag:bool-marking}
    \begin{tikzpicture}[diagram,baseline=(U-l.base)]
      \node (Elim) {$L$};
      \node (U/bool) [below = of Elim] {$\IHom{\floors{\El}\brk{\floors{\TyBool}}}{\floors{\TY}}$};
      \node (E-r) [right = 4cm of Elim] {$\IHom{\floors{\ObjTerm}}{\floors{\EL}}$};
      \node (E-l) [left  = 4cm of Elim] {$\IHom{\floors{\ObjTerm}}{\floors{\EL}}$};
      \node (U-r) [right = 4cm of U/bool] {$\IHom{\floors{\ObjTerm}}{\floors{\TY}}$};
      \node (U-l) [left  = 4cm of U/bool] {$\IHom{\floors{\ObjTerm}}{\floors{\TY}}$};
      \node (E/bool) [above = of Elim] {$\IHom{\floors{\El}\brk{\floors{\TyBool}}}{\floors{\EL}}$};
      \path[->,exists] (Elim) edge (U/bool);
      \path[->] (E-l) edge node [left] {$\floors{\El}_!$} (U-l);
      \path[->] (E-r) edge node [right] {$\floors{\El}_!$} (U-r);
      \path[->] (U/bool) edge node [below] {$\floors{\TmTrue}^*$} (U-l);
      \path[->] (U/bool) edge node [below] {$\floors{\TmFalse}^*$} (U-r);
      \path[->,exists] (Elim) edge (E-l);
      \path[->,exists] (Elim) edge (E-r);
      \path[->] (E/bool) edge node [sloped,above] {$\floors{\TmTrue}^*$} (E-l);
      \path[->] (E/bool) edge node [sloped,above] {$\floors{\TmFalse}^*$} (E-r);
      \path[->,exists] (E/bool) edge node [upright desc] {$\exists!$} (Elim);
      \path[->,bend right=50] (E/bool) edge node [upright desc] {$\floors{\El}_!$} (U/bool);
    \end{tikzpicture}
  \end{equation}

  In \cref{diag:bool-marking} above, we intend the cone under $L$ to be
  limiting. Finally, we define $\Mor[\phi]{\Ar{d}}{\SketchCat}$ to be the
  subdiagram of \cref{diag:bool-generators:1} omitting the section $s$.
\end{node}

\begin{node}
  The main subtlety of \cref{node:sketch-bool-elim} is the discrepancy between
  $\IHom{\floors{\ObjTerm}}{\floors{X}}$ and $\floors{X}$; these are not
  isomorphic at ``sketch-time'', but they will ultimately be canonically
  isomorphic in any model of the sketch because we have elsewhere marked
  $\floors{\ObjTerm}$ as the terminal object. Hence any model of our sketch
  will support an eliminator of the form exposed in
  \cref{node:bool-elim-intuition}.
\end{node}

\begin{node}
  The \emph{strict} booleans can be added to our theory by making $s$ an
  isomorphism; of course, these booleans will not be realized as a true
  colimit, but they will \emph{appear} to be so from the perspective of
  anything classified by $\TY$.
\end{node}

\subsection{Alternatives to sketching}

\CounterZeroNext{node}

\begin{node}
  While we have emphasized the 2-monadic perspective on LCCCs, there are several alternative ways of
  presenting locally Cartesian closed categories.
\end{node}

\begin{node}
  Recently \citet{bidlingmaier:2020} used the machinery of model categories to isolate locally
  Cartesian closed categories as the fibrant objects of a model category. The fibrant replacement
  operation then gives a free LCCC over a given sketch of a type theory.
  This method also endows $\LCCC$ with homotopical structure which, for instance, allows one to
  calculate the pushout of type theories by forming the \emph{homotopy pushout} of their sketches.
\end{node}

\begin{node}
  One may also use the (exhaustive) proof of the biequivalence between locally Cartesian closed
  categories and extensional Martin-L{\"o}f type theory~\citep{castellan-clairambault-dybjer:2017}.
  The free LCCC over some set of generators can be formed by taking the syntactic model of
  extensional type theory generated by this set of constants. This technique allows us to define a
  type theory by something akin to a signature in a logical framework.
\end{node}

\begin{node}
  From here on out, we take for granted that such objects can be constructed in
  one of several well-understood and equivalent ways, depending on the
  preferences of the reader for semantic vs.\ syntactic methods.
\end{node}

\section{Adequacy of LCCCs over representable map categories}
\label{sec:conservative}

\begin{node}
  Let $\Sigma$ be the signature of a type theory in Uemura's logical
  framework~\citep{uemura:2019}; this generates a syntactic representable map
  category $\RepThy$, but forgetting the difference between representable sorts
  and ordinary sorts, we also have a syntactic locally Cartesian closed
  category $\ECat$ using the results from \cref{sec:sketching}.
\end{node}

\begin{node}
  Regard $\ECat$ as a representable map category equipped with its maximal
  representable map structure; we have a representable map functor
  $\Mor[\ell]{\RepThy}{\ECat}$, \emph{i.e.}\ a lex functor that preserves pushforwards
  along representable maps. We will show that $\ell$ is full and
  faithful, using an argument explained by \citet{taylor:1999}.
\end{node}

\subsection{The gluing construction}

\begin{node}
  The structure map $\Mor[\Lift]{\RepThy}{\ECat}$ induces a left exact nerve
  $\Mor[\Nv]{\ECat}{\Psh{\RepThy}}$ like so:
  \[
    \Nv{E} : \Mor|{|->}|{X}{\Hom[\ECat]{\Lift{X}}{E}}
  \]
\end{node}

\begin{node}
  We form the Artin gluing of the nerve $\Mor[\Nv]{\ECat}{\Psh{\RepThy}}$ as follows:
  \[
    \DiagramSquare{
      nw = \GlCat,
      ne = \Psh{\RepThy},
      se = \Psh{\RepThy},
      east = \ArrId{\Psh{\RepThy}},
      south = \Nv,
      sw = \ECat,
      west = \GlFib,
      east/style = fibration,
      west/style = fibration,
      se/style = comma,
    }
  \]
\end{node}

\subsection{The Yoneda model}

\begin{node}
  We may define a functor $\Mor[\YoMod]{\RepThy}{\GlCat}$ taking each $X:\RepThy$ to
  the computability structure $\Mor{\Yo[\RepThy]{X}}{\Nv{\Lift{X}}}$ determined
  by the identity $\Mor{\Lift{X}}{\Lift{X}}$ under the following identification:
  \[
    \Hom[\Psh{\RepThy}]{\Yo[\RepThy]{X}}{\Nv{\Lift{X}}}
    \cong
    \Hom[\ECat]{\Lift{X}}{\Lift{X}}
  \]

  Recalling the adjunction $\Lift_!\dashv\Lift^*$ we note that $\Nv{\Lift{X}}
  \cong \Lift^*\Lift_!\Yo{X}$. Hence, we may give an explicit computation of
  $\YoMod{X}$ as either the unit
  $\Mor[\eta\Sub{\Yo{X}}]{\Yo{X}}{\Lift^*\Lift_!\Yo{X}}$ or the transpose
  $\Mor[\ArrId{\Lift_!\Yo{X}}\Transp]{\Yo{X}}{\Lift^*\Lift_!\Yo{X}}$.
\end{node}

\begin{node}\label{node:m-rep-map-fun-2}
  \emph{The functor $\YoMod$ is a model of $\RepThy$.}
  Clearly $\YoMod$ preserves finite limits, as these are determined ``pointwise'' in
  $\GlCat$ and finite limits are preserved by the Yoneda embedding and
  $\Mor[\Lift]{\RepThy}{\ECat}$; it remains to show that $\YoMod$ preserves
  pushforwards along representable maps. We fix a representable map
  $\FibMor[f]{X}{Y}$ and a map $\Mor[g]{Z}{X}$, to show that
  $\Mor[\YoMod{f_*g}]{\Yo{f_*g}}{\Nv{\Lift{f_*g}}}$ is the pushforward of $\YoMod{g}$
  along $\YoMod{f}$. Accordingly, we will exhibit the following bijection natural in
  a family $\Mor[p]{A}{Y}$ in $\GlCat$:
  \begin{equation}
    \Hom[\Sl{\GlCat}{\YoMod{X}}]{p^*\YoMod{f}}{\YoMod{g}}
    \cong \Hom[\Sl{\GlCat}{\YoMod{Y}}]{p}{\YoMod{f_*g}}
    \label[identity]{iso:gl-pi-transpose}
  \end{equation}

  We may assume without loss of generality that the domain of $A$ is
  representable, so we have $\Mor[A]{\Yo{\tilde{A}}}{\Nv{\GlFib{A}}}$.
  We first construct a map from the right-hand side to the left-hand side of
  \cref{iso:gl-pi-transpose}, fixing a morphism
  $\Mor[\alpha]{p}{\YoMod{f_*g}}:\Sl{\GlCat}{\YoMod{Y}}$ that we unfold like so:
  \begin{equation}
    \label[diagram]{diag:map-into-gl-pi:0}
    \begin{tikzpicture}[diagram,baseline=(sw.base)]
      \SpliceDiagramSquare{
        nw = \Yo{\tilde{A}},
        sw = \Nv{\GlFib{A}},
        ne = \Yo{f_*g},
        se = \Nv{\Lift{f_*g}},
        width = 4cm,
        height = 4cm,
        west = A,
        east = \YoMod{f_*g},
        south = \Nv{\GlFib{\alpha}},
        north = \tilde{\alpha},
      }
      \node (MY/tot) [color=gray,between = nw and ne, yshift = -1cm] {$\Yo{Y}$};
      \node (MY/base) [color=gray,between = sw and se, yshift = 1cm] {$\Nv{\Lift{Y}}$};
      \path[->,color=gray] (MY/tot) edge node [upright desc] {$\YoMod{Y}$} (MY/base);
      \path[->,color=gray] (nw) edge (MY/tot);
      \path[->,color=gray] (ne) edge (MY/tot);
      \path[->,color=gray] (sw) edge (MY/base);
      \path[->,color=gray] (se) edge (MY/base);
    \end{tikzpicture}
  \end{equation}

  Both the Yoneda embedding and $\Lift$ preserve pushforwards along
  representable maps, so $\Nv{\Lift{f_*g}} = \Nv{\Lift{f}_*\Lift{g}}$ and
  $\Yo{f_*g} = \Yo{f}_*\Yo{g}$. Accordingly we may transpose $\GlFib{\alpha}$ and
  $\tilde{\alpha}$ to obtain commuting triangles in $\RepThy$ and $\ECat$
  respectively, writing $\lambda/\bar\lambda$ for the transposition isomorphism
  of the pushforward:
  \begin{equation}
    \label[diagram]{diag:map-into-gl-pi:1}
    \begin{tikzpicture}[diagram,baseline=(X0.base)]
      \node (PB1) {$\tilde{A} \times_Y X$};
      \node [right = 6cm of PB1] (PB0) {$\GlFib{A} \times_{\Lift{Y}} \Lift{X}$};
      \node [right = 3cm of PB1] (Z1) {$Z$};
      \node [right = 3.5cm of PB0] (Z0) {$\Nv{Z}$};

      \node[between = PB1 and Z1] (B1) {};
      \node[between = PB0 and Z0] (B0) {};

      \node[below = of B1] (X1) {$\Yo{X}$};
      \node[below = of B0] (X0) {$\Lift{X}$};

      \path[->] (PB1) edge node[above] {$\bar\lambda\prn{\tilde{\alpha}}$} (Z1);
      \path[->] (PB0) edge node[above] {$\bar\lambda\prn{\GlFib{\alpha}}$} (Z0);
      \path[->] (PB1) edge (X1);
      \path[->] (PB0) edge (X0);
      \path[->] (Z1) edge (X1);
      \path[->] (Z0) edge (X0);
    \end{tikzpicture}
  \end{equation}

  It remains to prove that these diagrams assemble into a diagram of the following form, writing
  $-^\sharp$ for the transpose of the adjunction $\Lift_!\dashv\Lift^*$:
  \begin{equation}
    \DiagramSquare{
      nw = \Yo{\tilde{A}\times_{Y} X},
      ne = \Yo{Z},
      north = \Yo{\bar\lambda\prn{\tilde{\alpha}}},
      south = \Nv{\bar\lambda\prn{\GlFib{\alpha}}},
      east = \ArrId{\Lift{Z}}\Transp,
      west = \prn{A,\ArrId{\Lift{X}}}\Transp,
      sw = \Nv{\GlFib{A}\times_{\Lift{Y}}\Lift{X}},
      se = \Nv{\Lift{Z}},
      width = 5cm,
      height = 2.5cm,
    }
  \end{equation}

  Unfolding, this is equivalent to proving $\bar\lambda\prn{\GlFib{\alpha}} \circ \prn{A,
  \ArrId{\Lift{X}}} = \bar\lambda\prn{\tilde{\alpha}}$ as morphisms
  $\Mor{\Lift{\tilde{A}} \times_{\Lift{Y}} \Lift{X}}{\Lift{Z}}$. By the
  naturality of tranposition, however, this is precisely equivalent to
  $\GlFib{\alpha} \circ A = \tilde{\alpha}$, the content of the square in
  \cref{diag:map-into-gl-pi:0}. Accordingly the bijective map sending
  $\prn{\GlFib{\alpha},\tilde{\alpha}}$ to $\prn{\bar\lambda\prn{\GlFib{\alpha}},\bar\lambda\prn{\tilde{\alpha}}}$
  restricts to a map between the appropriate hom sets in $\GlCat$.
  A similar argument shows that the inverse to the transposition restricts to a map of the following type:
  \[
    \Mor{\Hom[\Sl{\GlCat}{\YoMod{X}}]{p^*\YoMod{f}}{\YoMod{g}}}{\Hom[\Sl{\GlCat}{\YoMod{Y}}]{p}{\YoMod{f_*g}}}
  \]

  Therefore \cref{iso:gl-pi-transpose} holds and hence $\YoMod$ preserves dependent products along
  representable map products.
\end{node}

\begin{node}
  Moreover, $\Mor[\YoMod]{\RepThy}{\GlCat}$ is fully faithful; to check fullness, we
  fix a morphism $\Mor[f]{\YoMod{X}}{\YoMod{Y}}$, \emph{i.e.}\ a commuting square in
  $\Psh{\RepThy}$ configured like so:
  \begin{equation}\label[diagram]{diag:M-ff:0}
    \DiagramSquare{
      nw = \Yo[\RepThy]{X},
      ne = \Yo[\RepThy]{Y},
      sw = \Nv{\Lift{X}},
      se = \Nv{\Lift{Y}},
      south = \Nv{\GlFib{f}},
      west = \YoMod{X},
      east = \YoMod{Y},
      north = \Yo[\RepThy]{\tilde{f}},
      width = 4.5cm,
    }
  \end{equation}

  It remains to check that
  $\YoMod{\tilde{f}}$ is the above square, which is the same as to show that
  $\Lift{\tilde{f}} = \GlFib{f}$; but this is exactly the content of \cref{diag:M-ff:0}
  instantiated at $\ArrId{X}$, hence $\YoMod$ is full. Faithfulness is clear,
  because the underlying $\RepThy$-map can be projected from any such square.
\end{node}

\subsection{The conservativity result}

\begin{node}
  We may now show that $\Mor[\Lift]{\RepThy}{\ECat}$ is fully faithful.
  First we
  observe that there is a locally Cartesian closed functor
  $\Mor[\YoModExt]{\ECat}{\GlCat}$ extending the representable map functor
  $\Mor[\YoMod]{\RepThy}{\GlCat}$ in the following configuration:
  \[
    \begin{tikzpicture}[diagram]
      \SpliceDiagramSquare{
        nw = \GlCat,
        ne = \Psh{\RepThy},
        se = \Psh{\RepThy},
        east = \ArrId{\Psh{\RepThy}},
        south = \Nv,
        sw = \ECat,
        west = \GlFib,
        east/style = fibration,
        west/style = fibration,
        se/style = comma,
        west/node/style = upright desc,
      }
      \node (RepThy) [left = of nw] {$\RepThy$};
      \node (Thy) [above = of nw] {$\ECat$};
      \path[->] (RepThy) edge node [above,sloped] {$\Lift$} (Thy);
      \path[->] (RepThy) edge node [below,sloped] {$\Lift$} (sw);
      \path[embedding] (RepThy) edge node [upright desc] {$\YoMod$} (nw);
      \path[->] (Thy) edge node [upright desc] {$\YoModExt$} (nw);
    \end{tikzpicture}
  \]
\end{node}

\begin{node}
  Because the gluing fibration $\FibMor[\GlFib]{\GlCat}{\ECat}$ preserves all
  locally Cartesian closed structure, we have a canonical isomorphism
  $\GlFib\circ\YoModExt \cong \ArrId{\ECat}$; because $\YoModExt$ is hence a section of the
  gluing fibration, it is faithful.
\end{node}

\begin{node}
  We observe that $\Mor[\Lift]{\RepThy}{\ECat}$ is faithful using the fact that
  $\YoModExt$ is faithful and $\YoMod=\YoModExt\circ\Lift$ is faithful. Fix $\Mor[f,g]{X}{Y}$ such
  that $\Lift{f} = \Lift{g}$; hence $\YoMod{f} = \YoMod{g}$ and because $\YoMod$ is
  faithful, we know that $f=g$.
  To see that $\Lift$ is full,
  we fix an arbitrary morphism
  $\Mor[f]{\Lift{X}}{\Lift{Y}}$; by functoriality we have a morphism
  $\Mor[\YoModExt{f}]{\YoMod{X}}{\YoMod{Y}}$, but $\YoMod$ is full so this must
  be the image of some $\Mor{X}{Y}$ under $\YoMod$.
\end{node}

\section*{Acknowledgment}

We thank Carlo Angiuli, Steve Awodey, Martin Bidlingmaier, Robert Harper, and Colin Zwanziger
for helpful discussions during the production of these notes.

This work was supported in part by AFOSR under grants MURI FA9550-15-1-0053 and
FA9550-19-1-0216.  Any opinions, findings and conclusions or recommendations
expressed in this material are those of the authors and do not necessarily
reflect the views of the AFOSR.

\nocite{kinoshita-power-takeyama:1999}
\nocite{uemura:2019}
\printbibliography

\end{document}